\documentclass[11pt]{elsarticle}
\usepackage{front,front_elsarticle_vd}

\usepackage{amsmath,amssymb,atmath}                      

\usepackage{pstricks,pst-node,pst-plot,pst_ext}          
\SpecialCoor
\usepackage{graphicx}

\usepackage{url,subfig}

\begin{document}

\newcommand{\mytitle}{Minimum-weight double-tree shortcutting for Metric TSP:
Bounding the approximation ratio}


\newcommand{\myabstract}{%
The Metric Traveling Salesman Problem (TSP) 
is a classical NP-hard optimization problem.
The double-tree shortcutting method for Metric TSP 
yields an exponentially-sized space of TSP tours,
each of which approximates the optimal solution 
within at most a factor of $2$.
We consider the problem of finding among these tours
the one that gives the closest approximation,
i.e.\ the \emph{minimum-weight double-tree shortcutting}.
Previously, we gave an efficient algorithm for this problem,
and carried out its experimental analysis.
In this paper, we address the related question 
of the worst-case approximation ratio 
for the minimum-weight double-tree shortcutting method.
In particular, we give lower bounds on the approximation ratio
in some specific metric spaces:
the ratio of 2 in the discrete shortest path metric,
1.622 in the planar Euclidean metric,
and 1.666 in the planar Minkowski metric.
The first of these lower bounds is tight;
we conjecture that the other two bounds are also tight,
and in particular that the minimum-weight double-tree method
provides a $1.622$-approximation for planar Euclidean TSP.}

\myfront

\section{Introduction}
\label{s-intro}

The Metric Travelling Salesman Problem (TSP) 
is a classical combinatorial optimization problem.
We represent a set of $n$ points in a metric space
by a complete weighted graph on $n$ nodes,
where the weight of an edge is defined 
by the distance between the corresponding points.
The objective of Metric TSP is to find in this graph 
a minimum-weight Hamiltonian cycle
(equivalently, a minimum-weight tour visiting every node at least once).
The most common example of Metric TSP is the planar Euclidean TSP, 
where the points lie in the two-dimensional Euclidean plane,
and the distances are measured according to the Euclidean metric.

Metric TSP, even restricted to planar Euclidean TSP, 
is well-known to be NP-hard \cite{Papadimitriou:77}.
Metric TSP is also known to be NP-hard to approximate 
to within a ratio $1.00456$,
but polynomial-time approximable to within a ratio $1.5$.
Fixed-dimension Euclidean TSP is known to have a PTAS
(i.e.\ a family of algorithms 
with approximation ratio arbitrarily close to $1$) \cite{Arora:98};
this generalises to any metric 
defined by a fixed-dimension Minkowski vector norm.

Two simple methods, 
double-tree shortcutting \cite{Rosenkrantz+:77}
and Christofides' \cite{Christofides:76,Serdyukov:78},
allow one to approximate the solution of Metric TSP 
within a factor of $2$ and $1.5$, respectively.
Both these methods belong to the class of \emph{tour-constructing heuristics},
i.e.\ ``heuristics that incrementally construct a tour 
and stop as soon as a valid tour is created'' \cite{Johnson_McGeoch:02}.
In both methods, we build an Eulerian graph on the given point set,
select an Euler tour of the graph,
and then perform \emph{shortcutting} on this tour
by removing repeated nodes, until all node repetitions are removed.
In general, it is not prescribed which one of several occurrences
of a particular node to remove.
Therefore, the methods yield an exponentially-sized space of TSP tours
(shortcuttings of a specific Euler tour in a specific Eulerian graph),
each of which approximates the optimal solution within 
at most a factor of $2$ (respectively, $1.5$).

The two methods differ in the way 
the initial weighted Eulerian graph is constructed.
Both start by finding the graph's minimum-weight spanning tree (MST).
The double-tree method then doubles every edge in the MST,
while the Christofides method adds to the MST a minimum-weight matching 
built on the set of odd-degree nodes\@.
The weight of the resulting Euler tour is higher than the optimal TSP tour
at most by a factor of $2$ (respectively, $1.5$),
and the subsequent shortcutting can only decrease the tour weight.

While any tour obtained by shortcutting of the original Euler tour
approximates the optimal solution 
within at most a factor of $2$ (respectively, $1.5$),
clearly, it is still desirable to find 
the shortcutting that gives the closest approximation.
Given an Eulerian graph on a set of points,
we will consider its \emph{minimum-weight shortcutting}
across all shortcuttings of all possible Euler tours of the graph.
We shall correspondingly speak 
about \emph{the minimum-weight double-tree}
and \emph{the minimum-weight Christofides} methods.

Unfortunately, for the general Metric TSP
(i.e.\ an arbitrary complete weighted graph with the triangle inequality),
the corresponding double-tree and Christofides 
minimum-weight shortcutting problems are both NP-hard.
The minimum-weight double-tree shortcutting problem
was also believed for a long time to be NP-hard for planar Euclidean TSP,
until a polynomial-time algorithm was given 
by Burkard et al.\ \cite{Burkard+:98}.
In \cite{Deineko_Tiskin:07}, we gave an improved algorithm
running in time $O(4^d n^2)$,
where $d$ is the maximum node degree in the rooted minimum spanning tree
(e.g.\ in the non-degenerate planar Euclidean case, $d \leq 4$).
In contrast, the Christofides version of the problem 
remains NP-hard even for planar Euclidean TSP
\cite{Papadimitriou_Vazirani:84}.

\begin{figure}
\psset{unit=0.7}

\newcommand{\dfdtree}{%
\multips(0,0)(1,0){6}{\psline{*-*}(0,1)}
\multips(0,1)(1,0){5}{\psline{*-}(0.2,0)(1,0)}}

\subfloat[Minimum spanning tree]{%
\centering \label{f-dtree-lb-mst}
\begin{pspicture}(0,-0.5)(5,1.5)
\dfdtree \psline(0,1)(5,1)
\scriptsize
\uput[l](0,0.5){$1$}
\uput[u](0.1,1){$\epsilon$}
\uput[u](0.6,1){$1$}
\end{pspicture}}
\hfill
\subfloat[Depth-first double-tree tour]{%
\centering \label{f-dtree-lb-output}
\begin{pspicture}(0,-0.5)(5,1)
\dfdtree \psline(0,1)(1,1)
\multips(1,0)(1,0){4}{\psline(0.2,1)}
\pcarc(5,0)(0,0)
\end{pspicture}}
%
\hfill
\subfloat[Absolute minimum-weight tour]{%
\centering \label{f-dtree-lb-minweight}
\begin{pspicture}(0,-0.5)(5,1)
\psframe[dimen=middle](0,0)(5,1)
\psset{linestyle=none} \dfdtree
\end{pspicture}}

\caption{\label{f-dtree-lb}
The depth-first double-tree method: a lower-bound instance}

\end{figure}
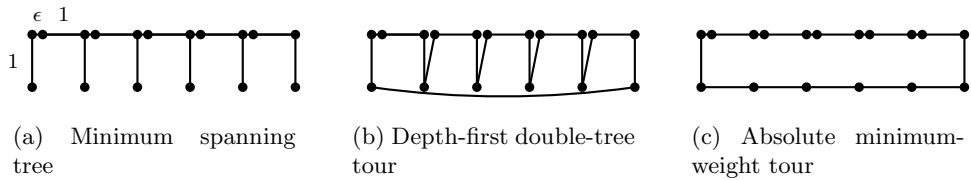

A natural question about the properties 
of the two approximation methods and their variants
is whether the approximation ratios $2$ and $1.5$ are tight,
i.e.\ whether there is a problem instance where the approximate solution 
has approximation ratio $2$ (respectively, $1.5$),
or a family of problem instances where the approximate solutions
approach these ratios arbitrarily closely.

For the minimum-weight double-tree method, 
the answer to this question is unknown,
as observed e.g.\ in \cite{Letchford_Pearson:08}.
The only existing lower bounds for the double-tree method 
apply to a shortcutting that is performed
in some suboptimal, easily computable order.
An example of such an order is depth-first tree traversal;
we shall call the resulting method \emph{depth-first double-tree shortcutting}.
A tight lower bound for this method is given 
by the standard Euclidean lower-bound construction 
shown in  \figref{f-dtree-lb}, 
whichadapted from \cite{Johnson_Papadimitriou:85}.
\figref{f-dtree-lb-mst} shows an instance point set 
and the (unique) minimum spanning tree.
We assume that $\epsilon=o(1)$; for example, we can take $\epsilon=1/n$.
The vertical size of the instance set is $1$,
and the horizontal size is $\bigpa{1+o(1)}n$.
The weight of the unique MST is $\bigpa{2+o(1)}n$;
the double-tree weight is $\bigpa{4+o(1)}n$.
The double tree undergoes no significant shortcutting,
and the resulting tour (\figref{f-dtree-lb-output}) 
still has weight $\bigpa{4+o(1)}n$.
The absolute minimum-weight tour (\figref{f-dtree-lb-minweight})
has weight $\bigpa{2+o(1)}n$,
therefore the approximation ratio on the given instance set is $2$.

\begin{figure}
\psset{unit=0.7}

\newcommand{\christo}{%
\multips(0,0)(1,0){6}{\psline{*-*}(0,1)}
\multips(0,1)(2,0){3}{\psline(1,0) \psdot(0.2,0)}
\multips(1,0)(2,0){2}{\psline(1,0) \psdot(0.2,0)}}

\subfloat[Minimum spanning tree]{%
\centering \label{f-Christo-lb-mst}
\begin{pspicture}(0,-0.5)(5,1.5)
\christo
\scriptsize
\uput[l](0,0.5){$1$}
\uput[u](0.1,1){$\epsilon$}
\uput[u](0.6,1){$1$}
\end{pspicture}}
\hfill
\subfloat[Minimum-weight Christofides tour]{%
\centering \label{f-Christo-lb-output}
\begin{pspicture}(0,-0.5)(5,1)
\christo
\pcarc(5,0)(0,0)
\end{pspicture}}
%
\hfill
\subfloat[Absolute minimum-weight tour]{%
\centering \label{f-Christo-lb-minweight}
\begin{pspicture}(0,-0.5)(5,1)
\psframe[dimen=middle](0,0)(5,1)
\psset{linestyle=none} \christo
\end{pspicture}}

\caption{\label{f-Christo-lb}
The minimum-weight Christofides method: a lower-bound instance}

\end{figure}
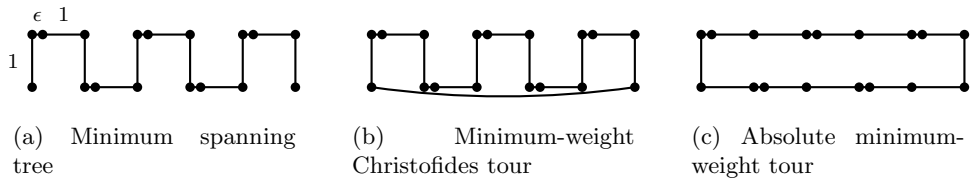

For the minimum-weight Christofides algorithm, a tight lower bound is given 
by the standard Euclidean lower-bound construction 
shown in \figref{f-Christo-lb}, 
which is adapted from \cite{Cornuejols_Nemhauser:78}
and uses the same conventions as \figref{f-dtree-lb}.
The minimum spanning tree has exactly two odd-degree nodes,
therefore the additional matching consists of a single edge.
The resulting Eulerian graph
(\figref{f-Christo-lb-output}) is already a Hamiltonian cycle,
hence no shortcutting is required.
The weight of the cycle is $\bigpa{3+o(1)}n$.
As before, the absolute minimum-weight tour (\figref{f-Christo-lb-minweight})
has weight $\bigpa{2+o(1)}n$,
therefore the approximation ratio on the given instance sets is $1.5$.

In the rest of this paper, we address the question 
of the worst-case approximation ratio 
for the minimum-weight double-tree shortcutting method
in some specific metric spaces.
In particular, we give a lower bound on the approximation ratio
in the discrete shortest path metric%
\footnote{The same result has been obtained independently
by Bil\`o at al.\ \cite{Bilo+:08}.};
this bound is tight, and can be regarded 
as a lower bound for a generic metric space.
We also give the first non-trivial lower bound
for the planar Euclidean and planar Minkowski metrics.

\section{The lower bounds}
\label{s-lbounds}

\subsection{The discrete shortest path metric}

The worst-case approximation ratio 
of the double-tree and Christofides methods
can clearly be dependent on the metric
in which the TSP problem is defined.
In the Introduction, we described a tight lower bound of $2$
on the worst-case approximation ratio in the planar Euclidean metric,
both for the depth-first version of the double-tree method
and for the minimum-weight Christofides method.
In contrast, no non-trivial lower bounds have been known,
to our knowledge, for the minimum-weight double-tree method in any metric.
A tight bound in the Euclidean metric seems difficult to obtain;
however, it can be established that the upper bound of $2$ 
is tight in some non-Euclidean metrics, 
and therefore is tight for the generic Metric TSP\@.

\newcommand{\DSPtree}[1]{%
\dotnode(0,0){r#1}
\dotnode(1,2){x#1 0}
\dotnode(2,1){x#1 1}
\dotnode(3,2){x#1 2}
\dotnode(4,0){x#1 3}
\dotnode(5,2){x#1 4}
\dotnode(6,1){x#1 5}
\dotnode(7,2){x#1 6}}

\newenvironment{DSPpic}{%
\begin{pspicture}(0,-0.5)(7,5)
\rput(0, 5){\psset{yunit=-1} \DSPtree{0}}
\rput(0, 0){\DSPtree{1}}
}{%
\end{pspicture}}

\begin{figure}
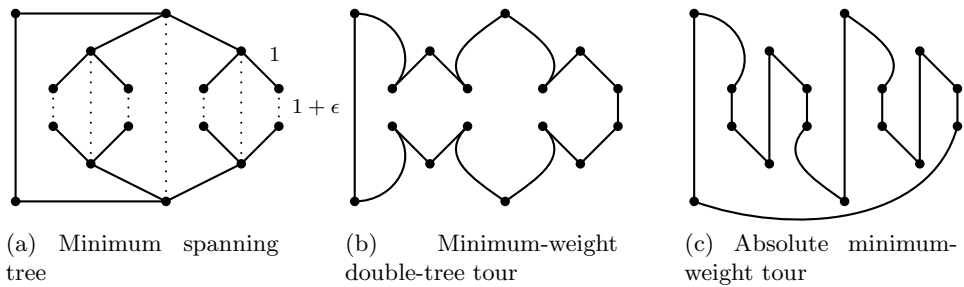

\psset{unit=0.5}

\subfloat[Minimum spanning tree]{%
\centering \label{f-DSP-lb-mst}
\begin{DSPpic}
\rput(x06){%
  \uput[ur](-0.5,0.5){\scriptsize $1$} 
  \uput[r](0,-0.5){\scriptsize $1+\epsilon$}
}
\multido{\Ik=0+1}{2}{%
  \psline(r\Ik)(x\Ik 3)
  \psline(x\Ik 0)(x\Ik 1)%
  \psline(x\Ik 1)(x\Ik 2)%
  \psline(x\Ik 1)(x\Ik 3)%
  \psline(x\Ik 3)(x\Ik 5)%
  \psline(x\Ik 4)(x\Ik 5)%
  \psline(x\Ik 5)(x\Ik 6)%
}
\multido{\Il=0+1}{7}{%
  \psline[linestyle=dotted](x0\Il)(x1\Il)
}
\psline(r0)(r1)
\end{DSPpic}}
\hfill
\subfloat[Minimum-weight double-tree tour]{%
\centering \label{f-DSP-lb-output}
\begin{DSPpic}
\psset{ncurv=1}
\pcline(r0)(r1)
\pccurve[angleA=0,angleB= 45](r0)(x00)
\pccurve[angleA=0,angleB=-45](r1)(x10)
\pcline(x06)(x16)
%
\pcline(x00)(x01)
\pcline(x01)(x02)
\pccurve[angleA=135,angleB=-145](x02)(x03)
\pccurve[angleA=-35,angleB=  45](x03)(x04)
\pcline(x04)(x05)
\pcline(x05)(x06)
\pcline(x10)(x11)
\pcline(x11)(x12)
\pccurve[angleA=-135,angleB=145](x12)(x13)
\pccurve[angleA=  35,angleB=-45](x13)(x14)
\pcline(x14)(x15)
\pcline(x15)(x16)
\end{DSPpic}}
%
\hfill
\subfloat[Absolute minimum-weight tour]{%
\centering \label{f-DSP-lb-minweight}
\begin{DSPpic}
\psset{ncurv=1}
\pcline(r0)(r1)
\pccurve[angleA=0,angleB=45](r0)(x00)
\pcline(x00)(x10)
\pcline(x10)(x11)
\pcline(x11)(x01)
\pcline(x01)(x02)
\pcline(x02)(x12)
\pccurve[angleA=-135,angleB=145](x12)(x13)
\pcline(x13)(x03)
\pccurve[angleA= -35,angleB= 45](x03)(x04)
\pcline(x04)(x14)
\pcline(x14)(x15)
\pcline(x15)(x05)
\pcline(x05)(x06)
\pcline(x06)(x16)
\pccurve[angleA=-105,angleB=-20,ncurv=0.75](x16)(r1)
\end{DSPpic}}

\caption{\label{f-DSP-lb}
The minimum-weight double-tree method: 
a lower-bound instance in the discrete shortest path metric}

\end{figure}

Given a weighted undirected graph,
consider the \emph{discrete shortest path metric} on its node set.
The distance between two nodes in this metric is defined
as the weight of the shortest path connecting them in the graph.
Let $n$ be a power of $2$.
Let $T_n$ be a rooted tree on $n$ nodes,
where the root has a single child, 
which branches off into a complete binary tree with $n/2$ leaves.
We construct an instance graph on $2n$ nodes as follows.
First, we create two copies of the tree $T_n$,
keeping track of corresponding pairs of nodes 
(i.e.\ pairs of nodes which are copies of the same node in $T_n$).
We then give all the edges in each tree weight $1$,
and connect the roots of the two trees by a \emph{root edge} of weight $1$.
Finally, we connect every pair of corresponding non-root nodes in both trees
by a \emph{cross-edge} of weight $1+\epsilon$.
We assume that $\epsilon=o(1)$; for example, we can take $\epsilon=1/n$.
The instance graph corresponding to $n=8$
is shown in \figref{f-DSP-lb-mst},
where edges of weight $1$ and $1+\epsilon$ are represented,
respectively, by solid and dotted lines. 

The unique MST consists of both copies of $T_n$ plus the root edge,
and has weight $\bigpa{2-o(1)}n$;
the double-tree weight is $\bigpa{4-o(1)}n$.
Note that for any two nodes $a$, $b$ within the same copy of $T_n$,
the distance between $a$ and $b$ is equal 
to the weight of the path connecting these nodes in the tree.
Hence, a shortcutting from $a,b,c$ to $a,c$ 
can reduce the tour weight,
only if $a$ and $c$ belong to different copies of $T_n$.
Also note that any double-tree Euler tour of $T_n$ has weight $2n-2$.
Any Hamiltonian cycle of the complete weighted graph 
obtained by shortcutting the double-tree Euler tour
will contain a Hamiltonian path 
in a complete weighted subgraph induced by each copy of $T_n$.
The weight of a such a Hamiltonian path 
can differ from the weight of the double-tree tour of $T_n$
by at most the weight of a single edge, 
which cannot exceed $2 \log n = o(n)$.
Therefore, the resulting Hamiltonian cycle 
still has weight $\bigpa{4-o(1)} n$.

The minimum-weight double-tree tour for our example 
is shown in \figref{f-DSP-lb-output},
where straight edges have weights $1$ and $1+\epsilon$,
and curved edges have integer weights greater than $1$.
An edge's curvature indicates the layout 
of the shortest path along which the edge weight is measured.
The absolute minimum-weight tour has weight $\bigpa{2+o(1)} n$,
and consists of the root edge and all the cross-edges,
linked together by edge-disjoint paths in the two trees.
The absolute minimum-weight tour for our example 
is shown in \figref{f-DSP-lb-minweight},
using the same graphic conventions as in \figref{f-DSP-lb-output}.
The approximation ratio of the minimum-weight double-tree method
on the given instance set is $4/2=2$,
which matches the generic upper bound%
\footnote{Many variations on the described construction are possible.
We have chosen a variant that is easy to visualise.}.

\subsection{Euclidean and Minkowski metrics}

\newenvironment{EucMinkpic}{%
\begin{pspicture}(-3.5,-2.7)(3.5,3.5)
\dotnode(0;0){x}%
\multido{\Ik=0+1,\Ia=30+120}{3}{%
  \dotnode(0.5;\Ia){x\Ik}%
  \rput{\Ia}(1;\Ia){%
    \dotnode(0;0){x\Ik o}
    \multido{\Id=1+1}{3}{\dotnode(\Id;060){x\Ik l}}%
    \multido{\Id=1+1}{3}{\dotnode(\Id;300){x\Ik r}}%
}}}{%
\end{pspicture}}

\begin{figure}[tb]
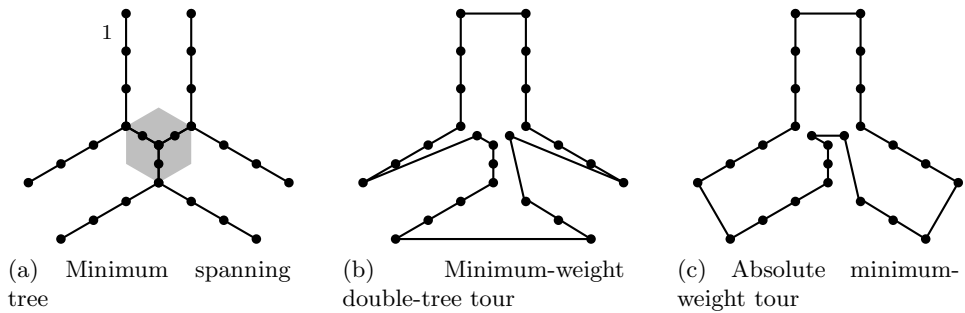

\psset{unit=0.5}

\subfloat[Minimum spanning tree]{%
\centering \label{f-EucMink-lb-mst}
\begin{EucMinkpic}
\pspolygon*[linecolor=gray25](1;30)(1;90)(1;150)(1;210)(1;270)(1;330)
\rput(x1r){ \uput[l](0,-0.5){\scriptsize $1$} }
\multido{\Ik=0+1}{3}{%
  \psdots(x)(x\Ik)(x\Ik o)      
  \psline(x)(x\Ik)(x\Ik o)(x\Ik l)%
  \psline(x)(x\Ik)(x\Ik o)(x\Ik r)%
}
\end{EucMinkpic}}
\hfill
\subfloat[Minimum-weight double-tree tour]{%
\centering \label{f-EucMink-lb-output}
\begin{EucMinkpic}
\pspolygon(x0l)(x1r)(x1o)(x1l)(x1)(x)(x2)(x2o)(x2r)(x2l)%
  (1.622;300)(x0)(x0r)(x0o)
\end{EucMinkpic}}
%
\hfill
\subfloat[Absolute minimum-weight tour]{%
\centering \label{f-EucMink-lb-minweight}
\begin{EucMinkpic}
\pspolygon(x0l)(x1r)(x1o)(x1l)(x2r)(x2o)(x2)(x)(x1)(x0)%
  (1.622;300)(x2l)(x0r)(x0o)
\end{EucMinkpic}}

\caption{\label{f-EucMink-lb}
The minimum-weight double-tree method: 
a lower-bound instance in the Euclidean and Minkowski metrics}

\end{figure}

Compared with the above construction for the discrete shortest path metric,
it appears to be much more difficult 
to obtain a tight bound in planar Euclidean-type metrics.
We describe a construction that provides 
the first non-trivial lower bound
on the approximation ratio of the minimum-weight double-tree method 
in the planar Euclidean and Minkowski metrics.

The proposed construction consists of $6n+1$ points,
and is shown in \figref{f-EucMink-lb-mst} for $n=4$.
The instance point set consists of seven points forming
a symmetric three-way central star of arbitrary constant size,
and six rows of points extending from the star's ends
in three symmetric directions in steps of length $1$.
\figref{f-EucMink-lb-mst} shows the (unique) minimum spanning tree,
which has weight $\bigpa{6+o(1)}n$.
\figref{f-EucMink-lb-output} shows the minimum-weight double-tree tour,
which has weight $\bigpa{8+\surd 3+o(1)}n$.
\figref{f-EucMink-lb-minweight} shows the absolute minimum-weight tour,
which has weight $\bigpa{6+o(1)}n$.
The approximation ratio of the minimum-weight double-tree method
on the given instance set is $(8+\surd 3)/6 \approx 1.622$.
There remains a substantial gap between this lower bound 
and the generic upper bound of $2$, which is also 
the best known upper bound in the planar Euclidean metric.

The same construction provides a somewhat stronger lower bound
in a metric defined by \emph{the hexagonal norm} ---
a Minkowski vector norm with the unit disc
in the shape of a regular hexagon (see \figref{f-EucMink-lb-mst}).
In this metric, the distance between two points
is measured along a polygonal path
composed from segments parallel to the edges of the unit disc.
The weights of the minimum spanning tree (\figref{f-EucMink-lb-mst}) 
and of the absolute minimum-weight tour (\figref{f-EucMink-lb-minweight})
on the above instance set remain asymptotically unchanged in the new metric.
However, the weight of the minimum-weight double-tree tour 
(\figref{f-EucMink-lb-output}) increases to $\bigpa{10+o(1)}n$.
Therefore, the lower bound in the hexagonal metric is $10/6 \approx 1.666$.

\section{Conclusions}
\label{s-concl}

In the previous section, we presented lower bounds 
on the minimum-weight double-tree method.
We have shown that the trivial upper bound of 2
is tight in at least some metrics 
(in particular, the discrete shortest path metric).
However, in the important cases of the Euclidean and Minkowski metrics,
a substantial gap remains between our lower bounds 
of 1.622 (respectively, 1.666) and the trivial upper bound of 2.
Considering the apparent difficulty of improving on these lower bounds,
and the good approximation behaviour of the minimum-weight double-tree algorithm
on typical Euclidean TSP instances \cite{Deineko_Tiskin:07},
we conjecture that these lower bounds are tight,
and that the minimum-weight double-tree method
provides a $1.622$-approximation for planar Euclidean TSP.

\bibliographystyle{plain}
\bibliography{auto,opt,tsp,books}

\end{document}